\documentclass[conference]{IEEEtran}
\IEEEoverridecommandlockouts
\usepackage{cite}
\usepackage{amsmath,amssymb,amsfonts}
\usepackage[T1]{fontenc}
\usepackage[utf8]{inputenc}
\usepackage{graphicx}
\usepackage{textcomp}
\usepackage{xcolor}
\usepackage{tabularx,array,booktabs,url,placeins}

\begin{document}
\title{What Is Under the Hood? A Cross-Ecosystem Study of Vulnerable Dependencies in Construction Software}
\author{\IEEEauthorblockN{Mahzabin Tamanna\IEEEauthorrefmark{1}, S M Jamil Uddin\IEEEauthorrefmark{2}}\IEEEauthorblockA{\IEEEauthorrefmark{1}Department of Computer Science, North Carolina State University, Raleigh, North Carolina, USA}\IEEEauthorblockA{\IEEEauthorrefmark{2}Department of Construction Management, Colorado State University, Fort Collins, Colorado, USA}}
\maketitle
\begin{abstract} 
The construction industry increasingly relies on software built from third-party components, yet the dependency composition and known-vulnerability exposure of construction-related software remain poorly analyzed. In this study, we conduct an empirical, cross-ecosystem investigation of 904 construction-related GitHub repositories spanning six construction functions and a multifunction category. Using commit-pinned repository snapshots and OSV-Scanner, we examine dependency assessability, known-vulnerability prevalence, severity, weakness classifications, and shared upstream exposure. Our results reveal that only 494 repositories (54.6\%) completed vulnerability assessment, revealing substantial gaps in dependency visibility. Among successfully scanned repositories, 395 (80.0\%) contained at least one dependency version associated with a published vulnerability advisory, and 97.2\% of affected repositories contained a \textit{High} or \textit{Critical} severity risk. Additionally, Common Weakness Enumeration (CWE) mapping identified recurring resource-consumption, path-traversal, information-exposure, and code-injection weaknesses. Vulnerability burden was uneven, with a median of 50 distinct advisory clusters per affected repository, while just two affected upstream package families collectively reached \textasciitilde{70}\% of affected repositories. These findings establish an empirical baseline for dependency assessability and known-vulnerability exposure in construction-related open-source software. Our findings underscore the importance of documenting assessment coverage, interpreting advisory matches in their deployment context, and examining shared upstream dependencies when evaluating construction software supply-chain security.\end{abstract}


\section{Introduction}

Digital construction increasingly depends on software built upon open source third-party components. Building information modeling (BIM) and model-checking tools, project-management platforms, digital twins, reality-capture pipelines, and connected site equipment increasingly shape how construction information is created, exchanged, and acted upon \cite{ref33,ref42}. As these tools move from prototypes and research code into routine project delivery, their security and maintainability depend not only on visible functionality but also on software security and assurance practices.

That limited visibility matters because an application can inherit vulnerabilities through third-party components used for functions such as file processing, web interfaces, authentication, and machine learning. These components may themselves depend on additional packages. Software Composition Analysis (SCA) makes dependency structure observable by identifying components and cross-referencing discovered components against public security databases like the National Vulnerability Database (NVD) and Common Vulnerabilities and Exposures (CVE) lists to flag security risks \cite{ref38}. Applied to construction-related software, SCA can indicate whether the assessed inputs identify a component version associated with a known vulnerability. Studies of general open-source ecosystems demonstrate that vulnerable dependencies can recur across many projects and persist in downstream applications \cite{ref12,ref39,ref51}. Whether those findings translate into operational exposure depends on how the components are built, deployed, and used.

The relevance of this software-layer problem to construction is evident via cyber incidents affecting construction organizations and advisories affecting components used in construction software. The UK Information Commissioner’s Office financially penalized the construction company Interserve \pounds{}4.4 million in October 2022 after a phishing-led intrusion compromised the systems and 16 accounts and encrypted personal data of up to 113,000 current and former employees, the regulator’s findings included failures to maintain supported systems and to act on vulnerability information \cite{ref23}. Bouygues reported a ransomware attack on 30 January 2020 that led to a precautionary shutdown of its information systems \cite{ref8}. Vendor disclosures show how exposure travels through shared components. Autodesk advisory ADSK-SA-2025-0024 lists 20 vulnerabilities in a shared component bundled with AutoCAD, Civil 3D, InfraWorks, Revit, Advance Steel, and other products, which could allow code execution if a user opens a maliciously crafted file \cite{ref3}. These cases show both that construction is exposed to cyber risk and why dependency-level analysis matters, as a single upstream weakness can create simultaneous exposure for many downstream users.

Shared software exposure is especially difficult to govern within construction’s project-based organization. Projects are delivered by temporary, multi-organizational teams of designers, contractors, specialist subcontractors, technology suppliers, and owners, and the firms involved are only loosely connected from one project to the next \cite{ref13}. Responsibility for selecting, updating, and operating a digital tool can pass across these interfaces or fall between them \cite{ref46,ref49}.

Construction cybersecurity research has examined many of these organizational and technological concerns. Studies have examined the state of cybersecurity research and practice through field-level analyses and scoping reviews \cite{ref26,ref41}, developed construction-specific risk and assessment frameworks \cite{ref46,ref49}, and investigated vulnerabilities in technologies and interactions such as BIM/common data environments and construction networks \cite{ref27}. This research shows growing recognition of cyber risk in construction, but its empirical focus has largely been on organizations, project participants, information environments, and individual construction technologies rather than on the software components from which those technologies are built.

In parallel, software-engineering research has shown that vulnerabilities in third-party and open-source dependencies can propagate widely through software ecosystems and persist in downstream applications \cite{ref12,ref39,ref51}. These studies, however, generally define their populations by programming language, package ecosystem, or broad collections of open-source projects rather than by application domain. To our knowledge, we found no empirical study that connects these two lines of work by systematically examining the dependency composition and known vulnerabilities of software developed for the construction sector. Consequently, it remains unclear whether construction software commonly incorporates vulnerable third-party components, whether particular dependencies create shared exposure across construction subdomains, and to what extent the software composition of such projects is observable from available dependency metadata. This lack of visibility is important because software-component transparency is a prerequisite for identifying and managing dependency-related vulnerabilities \cite{ref40,ref45}.

This study addresses the gap with a cross-ecosystem mining of software repositories, analyzing 904 construction-related GitHub repositories spanning BIM and model checking, project management, digital twins, reality capture, robotics and automation, and jobsite monitoring. We extracted dependency manifests and lockfiles from each repository at a fixed commit and scanned them with OSV-Scanner against the Open Source Vulnerabilities (OSV) database \cite{ref32}. The results are reported with explicit denominators for repositories that could not be assessed. The analysis is framed by loose-coupling theory \cite{ref31,ref47}, analyzing whether construction projects that are loosely coupled as organizations become tightly coupled through shared upstream software, and whether the composition of that software is visible enough to be governed.

The study pursued two goals: to establish an empirical baseline of known-vulnerable dependency exposure in construction-related open-source software, and to determine whether such exposure is sufficiently observable to support governance of construction software. Toward the goals, we studied five research questions (RQs) as follows:

RQ1: To what extent can construction-related repositories be assessed for vulnerable dependencies?

RQ2: How prevalent are vulnerable dependencies in construction projects? Does prevalence differ across subdomains and ecosystems?

RQ3: How severe are the vulnerabilities, and which security weaknesses characterize them?

RQ4: How much vulnerability burden do affected repositories carry, and how concentrated is exposure in shared packages?

RQ5: Do observable repository characteristics associate with vulnerable dependency exposure?

To our knowledge, this is the first construction-focused, cross-ecosystem repository study to quantify the assessability, prevalence, severity, shared-package concentration, and Common Weakness Enumeration (CWE) security weakness profile of known-vulnerable third-party dependencies in public architecture, engineering, and construction (AEC) software. The contribution is a domain-grounded measurement of construction software . Empirically, the study provides a construction-specific baseline with explicit accounting for repositories whose composition cannot be assessed. Theoretically, it extends loose-coupling theory to the software layer through the concepts of technical coupling and coupling asymmetry, and treats software-composition assessability as a precondition for cyber-risk governance in the construction engineering domain. Practically, it identifies where composition transparency should be required, which shared packages offer coordinated triage opportunities, and which types of weaknesses should guide construction-specific review questions.

\section{Literature Review}

\subsection{Construction Cybersecurity Research}

Construction cybersecurity has developed from general warnings into a recognizable research field. Mantha \& García de Soto \cite{ref26} used a bibliometric analysis to position construction relative to other sectors and to propose a research agenda. Salami Pargoo \& Ilbeigi \cite{ref41}  found in a scoping review that the literature concentrated on identifying risks and protecting assets, with less attention to other cybersecurity management functions. Earlier, Parn \& Edwards \cite{ref34} identified the common data environment that underpins BIM-enabled collaboration as a focal attack surface, and ISO 19650-5:2020 codified a security-minded approach to information management for built assets.

At the project and organizational levels, framework and risk-assessment studies examine how cyber risk spans construction activities, assets, and participants. Turk et al. \cite{ref46}  proposed a systemic framework that relates cybersecurity to information assets, material assets, people, and systems rather than to data confidentiality alone. Yao \& García de Soto \cite{ref49} developed a machine-learning-centric approach to assessing cyber risks in construction projects, and Yao et al. \cite{ref50} derived a taxonomy of project-level cyber risk factors, organized into five categories, from a large multi-source literature corpus. For digital twins, Evripidou et al. \cite{ref14} found that organizational challenges have received limited analysis and that research has emphasized confidentiality and privacy relative to other security dimensions. These studies establish that construction cyber risk is shaped by project structure and organizational responsibility. They characterize the risk through literature synthesis, expert judgment, and organizational analysis, however, rather than by measuring the software on which projects depend.

A smaller stream of work evaluates construction technology directly. Sonkor \& Soto \cite{ref42} reviewed operational technology on construction sites from a cybersecurity perspective, including autonomous and remotely operated equipment. Sonkor et al. \cite{ref43} then assessed the vulnerabilities of an autonomous site-monitoring system using the Common Vulnerability Scoring System (CVSS), with four assessors of different backgrounds. At a larger scale, Mantha et al. \cite{ref27} used an agent-based model to investigate the cyber vulnerabilities of construction networks, and Pärn et al. \cite{ref35}  synthesized digital-twin cybersecurity research around the cyber, physical, and social dimensions faced by general contracting firms. These studies move from describing threats to evaluating systems. Their units of analysis, however, are single systems or modeled networks. No single characteristic characterizes the components shared across the many tools that construction researchers, developers, and practitioners build, propagate, and reuse.

\subsection{Open-Source Software and Vulnerable Dependencies}

Software engineering research has established both the scale of the dependency problem and methods for measuring it. Decan et al. \cite{ref12}  analyzed how security vulnerabilities affect the npm package dependency network through direct and indirect dependencies, and Zerouali et al. \cite{ref51} extended this analysis to npm and RubyGems. Zimmermann et al. \cite{ref52} showed that individual packages and a small number of maintainer accounts could affect large parts of the npm ecosystem, a concentration that turns popular packages into single points of failure. At the project level, Prana et al. \cite{ref39}  studied 450 Java, Python, and Ruby projects over time, examining the distribution, severity, and persistence of vulnerable dependencies, and Alfadel et al. \cite{ref1} analyzed security vulnerabilities across Python packages.

\subsection{Software Transparency, Maintenance Practice, and Governance}

Research on countermeasures has focused on dependency-update automation. Alfadel et al. \cite{ref2}  found that most Dependabot security pull requests in the projects they studied were accepted, often within a day, which suggests that automation shortens exposure when maintainers are responsive. Policy has moved in parallel toward software transparency and supplier responsibility. US Executive Order 14028 directed federal software acquisition toward secure development practices and software bills of materials (SBOMs) \cite{ref53}, the National Telecommunications and Information Administration \cite{ref29} defined the minimum elements of an SBOM, and the NIST Secure Software Development Framework set expectations for identifying and remediating vulnerabilities throughout the software life cycle \cite{ref44}. The EU Cyber Resilience Act places cybersecurity obligations, including vulnerability handling, on those who place products with digital elements on the EU market. Its reporting obligations apply from 11 September 2026 and its remaining provisions from 11 December 2027 \cite{ref5}. None of this work, however, examines construction software as a population.

\subsection{Theoretical Lens: Loose Organizational Coupling and Tight Technical Coupling}

Loose-coupling theory provides a lens that connects these literatures. Weick \cite{ref47} described loosely coupled systems as those in which elements respond to one another while preserving their own identity and separateness. Such systems permit localized adaptation and can seal off a breakdown in one part from the rest, although the same looseness makes coordinated, system-wide change more difficult. Orton \& Weick \cite{ref31} reconceptualized loose coupling as a dialectic of distinctiveness and responsiveness and proposed researchers to specify which elements are coupled and through what mechanisms. Perrow \cite{ref37} supplied the contrasting logic of tight coupling: when processes are closely linked and leave little slack, a failure in one component propagates rapidly, before it can be isolated.

Construction management works have applied this vocabulary to the industry itself. Dubois \& Gadde \cite{ref13} characterized construction as combining tight couplings within individual projects with loose couplings in the permanent network of firms that persists across projects, and linked this pattern to the industry’s difficulties with productivity and innovation. Temporary multi-organizations form around each project and dissolve on completion \cite{ref9}, leaving few durable channels through which lessons, or responsibilities, carry from one project to the next.

Digital construction adds a technical layer to this organizational pattern. Tools built by different developers for different construction functions often reuse the same upstream packages. A vulnerability in an upstream package can therefore create shared exposure across independently developed tools that incorporate affected versions, spanning firms, projects, and construction subdomains. Here, technical coupling refers to dependence on a common upstream component, it does not establish that failures propagate directly between the applications. The organizational buffering that Weick \cite{ref47} attributed to loose coupling does not contain a weakness that arrives through shared code, while the looseness that Dubois \& Gadde \cite{ref13} associated with slow industry-wide learning is likely to slow coordinated remediation. We call this mismatch coupling asymmetry: exposure is tightly coupled across the sector, whereas the organizational capacity to respond is loosely coupled. Managing coupling asymmetry presupposes that the coupling can be seen. We therefore define software-composition assessability as the extent to which the components of a software artifact can be identified and matched against vulnerability knowledge with available tools.

\subsection{Research Gap}

Construction studies explain why cyber risk is systemic and organizationally shaped, but do not measure the software projects that they rely on. Studies of technical assessments evaluate individual systems or modeled networks and cannot reveal shared exposure. Software engineering studies measure dependency vulnerabilities , but in populations not defined by construction, and policy requires component-level transparency without an empirical baseline of what can be observed. The gap is therefore threefold. First, it is unknown how much of this software can be observed with standard tooling, which determines whether any exposure estimate is trustworthy and whether transparency requirements can be met (RQ1). Second, the prevalence, severity, weakness types, and concentration of known-vulnerable dependencies in construction-related software are unknown (RQ2-RQ4). Third, observable maintenance signals to exposure in this setting (RQ5). 

A construction-defined measurement is warranted even if the vulnerabilities are not unique to construction, because ecosystem mix, temporary multi-organizational delivery, and handover determine who must act on shared components and what evidence they can obtain. Its contribution lies in the population, the coverage accounting, and the theoretical framing, not in a new detection technique. Owners, contractors, and suppliers are being asked to provide component-level assurance \cite{ref29,ref44} without a baseline of which components recur across construction-related tools. Analyzing the state of dependencies in construction software is a critical gap this work studied. 

\section{Background and Definitions}

This section defines the software and security terms used in the rest of the paper.

\subsection{Packages, Ecosystems, and Dependencies}

A software package (or library) is a unit of reusable code that is distributed with a name and a version for inclusion in other software. A package registry is the public service that hosts packages and their metadata, and the set of packages in a registry, together with its users, tools, and conventions, is a package ecosystem (for example npm for JavaScript, PyPI for Python, Maven for Java, NuGet for .NET, and Packagist for PHP). Each ecosystem has its own dependency-declaration formats, version rules, and security-advisory practices, which is why this study reports results by ecosystem.

A dependency is a package that a piece of software needs in order to build or run. A direct dependency is one that the project declares itself. A transitive dependency is one that the project does not declare but that is required by one of its dependencies. The dependencies and the relationships between them form a dependency graph. For example, a PPE-detection tool might declare an image-processing package directly, that package may in turn require further packages that the tool’s developers never chose and may not know about. A vulnerable package version can therefore be included indirectly through another dependency. Whether its vulnerable functionality is present and reachable in a particular application requires further assessment \cite{ref12,ref52}.

Dependencies are also separated by scope. A runtime (production) dependency is needed when the software runs. A development dependency is needed only to build, test, or lint the software, such as a front-end build tool, and is normally not shipped to users. Vendored dependencies are copies of third-party source code stored inside the repository itself rather than fetched from a registry. Whether development-scope packages should count toward exposure matters for interpreting results \cite{ref36}.

\subsection{Declaring Dependencies: Manifests, Lockfiles, and Versions}

A manifest is a file in which a project declares its direct dependencies, usually with a version constraint that permits a range of versions (for example package.json, requirements.txt, pom.xml). A lockfile records the exact versions that were resolved, including transitive dependencies, so that a build is reproducible (for example, package-lock.json, poetry.lock, composer.lock, Cargo.lock). The distinction matters for measurement: a lockfile shows precisely which versions were installed, while a manifest may show only a range and only the direct dependencies, so a scan of a manifest must resolve versions and can see less of the graph.

A repository snapshot is the state of a repository at a specific commit, identified by a commit hash (SHA). Freezing repositories at a snapshot fixes the files being analyzed, but not necessarily the registry state used to resolve them.

\subsection{Software Composition Analysis and Software Bills of Materials}

Software composition analysis (SCA) identifies the third-party components in a software artifact and matches their versions against databases of known vulnerabilities \cite{ref38}. Its output is a list of findings, each linking a component version to an advisory. A software bill of materials (SBOM) is a formal inventory of the components in a piece of software, the minimum elements recommended for federal software supply-chain transparency are defined by NTIA \cite{ref29}. SCA and SBOMs address the same question, which is what a piece of software is made of, and both depend on the component list being obtainable in the first place.

\subsection{Vulnerabilities, Advisories, Severity, and Weaknesses}

A vulnerability is a weakness in software that an attacker could exploit to compromise confidentiality, integrity, or availability. A security advisory is a published record of a vulnerability that names the affected package, the affected and fixed versions, and one or more identifiers. Common identifiers include CVE (Common Vulnerabilities and Exposures), GHSA (GitHub Security Advisory), PYSEC (Python advisories), and RUSTSEC (Rust advisories). The Open-Source Vulnerabilities (OSV) database aggregates advisories from many ecosystems into one schema \cite{ref30,ref32}. Because one vulnerability is often published under several identifiers, records that list one another as aliases describe the same issue. In this study we merged such records into a single advisory cluster, more details are provided in Section IV.

The Common Vulnerability Scoring System (CVSS) assigns an advisory a base score from 0 to 10 that describes technical risk severity, which is  reported in the bands such as \textit{Low}, \textit{Medium}, \textit{High}, and \textit{Critical} \cite{ref15}. The Common Weakness Enumeration (CWE) is a catalog of weakness types, such as path traversal (CWE-22), that classifies why a vulnerability exists \cite{ref28}. CVSS therefore indicates how urgent an advisory is in security risk terms, and CWE indicates what kind of problem it is. A known-vulnerable dependency is a dependency or version of dependency that an advisory lists as affected. An exploitable vulnerability is one that an attacker can actually exploit in a given deployment. A reachable vulnerability is one in which the vulnerable code are reachable by the software. A scanner match establishes the first but not the other two. A vulnerability also has a lifecycle of introduction, discovery, public disclosure, and fix. The cross-ecosystem design here observes the state of dependencies at a single commit, not the timing of these events.

\section{Methods}

\subsection{Study Design and Data Collection}

We conducted a cross-ecosystem study by mining software repositories from public, construction-related GitHub repositories. The repository was the principal unit of analysis, while dependency inputs, package versions, advisory clusters, and repository-advisory pairs were secondary units used to describe burden without treating repeated occurrences of one advisory as independent vulnerabilities. A scanner match denotes a dependency version associated with an advisory in the OSV database at collection time \cite{ref38}. The query strings, screening rules, and aggregation choices are provided with the study to make the study reproducible. Our sampling decisions are reported following the guidelines of Baltes \& Ralph \cite{ref4}.

Candidate repositories were retrieved on 08/10/2026 through the GitHub REST Search API with the GitHub-API\cite{ref19}. A set of 26 queries was leveraged for data collection covering six construction functions: BIM/openBIM/model checking, construction project management, construction digital twins, reality capture/site sensing, construction robotics and automation, and jobsite IoT, monitoring, and safety. Search was restricted to public, non-fork repositories. We queried repository names, descriptions, and topics for relevant data. For each query, the collector retrieved a single page of up to 100 results, the maximum page size permitted by the GitHub API. No sort parameter was specified, so results were returned in GitHub's default best-match order \cite{ref18}. Across all queries, this procedure yielded 1,639 query-level records, which were reduced to 1,428 unique candidates after deduplication by repository identifier. The difference reflects repositories surfaced by more than one search query. No filtering was applied on minimum star count, repository age, activity level, or language. Because retrieval was governed by an undisclosed relevance-ranking algorithm rather than by uniform or random sampling, the resulting corpus consists of a relevance-ranked, non-probability sample of construction-related repositories rather than a census of construction software on GitHub \cite{ref4,ref25}.

Table~\ref{tab:1} summarizes the six construction functions covered by the 26 search queries. Search functions guided candidate retrieval. The final construction strata were assigned through repository screening. Repositories serving multiple central functions were classified as S07. Exact query strings are provided in Appendix A (Table A2), and collection artifacts are documented in Appendix E.

\begin{table*}[!t]
\caption{Search concepts and the number of GitHub queries per construction function.}
\label{tab:1}
\centering
\scriptsize
\setlength{\tabcolsep}{3pt}
\renewcommand{\arraystretch}{1.12}
\begin{tabularx}{\textwidth}{@{}>{\raggedright\arraybackslash}X>{\raggedright\arraybackslash}X>{\raggedright\arraybackslash}X>{\raggedright\arraybackslash}X@{}}
\toprule
\textbf{Construction function} & \textbf{Queries} & \textbf{Search concepts} & \textbf{Example exact query} \\
\midrule
BIM, openBIM, model checking & 5 (Q01-Q05) & Building information modeling, openBIM, BIM and IFC topics & topic:ifc fork:false is:public \\
Construction project management & 4 (Q06-Q09) & Construction management, project management, ERP & "construction project management" in:name,description,topics fork:false is:public \\
Construction digital twins & 3 (Q10-Q12) & Construction digital twin (phrase, combined, topic) & topic:construction-digital-twin fork:false is:public \\
Reality capture and site sensing & 4 (Q13-Q16) & Reality capture, photogrammetry, point cloud & construction photogrammetry in:name,description,topics fork:false is:public \\
Robotics and automation & 4 (Q17-Q20) & Construction robotics, robotic construction, construction robot & "construction robot" in:name,description,topics fork:false is:public \\
Jobsite IoT, monitoring, safety & 6 (Q21-Q26) & Smart construction site, site IoT, safety and computer vision, worker, hardhat and PPE detection & PPE detection construction in:name,description,topics fork:false is:public \\
\bottomrule
\end{tabularx}
\vspace{4pt}
\begin{minipage}{\textwidth}\scriptsize All queries were restricted to public, non-fork repositories. Search concepts summarize the intended coverage; example queries show one executed query per function. Repositories could appear in multiple query results and were deduplicated before eligibility screening.\end{minipage}
\end{table*}

\subsection{Eligibility Screening and Construction Strata Classification}

For this study data eligibility was assessed and filtered against inclusion and exclusion criteria. A repository was included if its description, documentation, topics, and content collectively presented direct support for construction or AEC work and if it contained meaningful executable source code. In this study, meaningful executable source code is defined as source files and project configuration sufficient to implement, run, process, control, or evaluate a construction-related software function. A dataset, model-weight archive, paper, tutorial, or link list without accompanying application source did not satisfy this criterion. Exclusion criteria covered repositories without construction relevancy, generic software lacking construction-specific implementation, documentations, dataset or model-only repositories, educational assignments and trivial demonstrations, duplicates or mirrors, unavailable or empty repositories, offensive tooling and repositories with readme only. To verify data quality, two authors independently assessed the relevance of all 1,428 candidate repositories to the construction industry using these criteria. The initial inter-rater agreement was  95.94\% (Cohen’s $\kappa$ = 0.91), and conflicts were resolved via a collaborative discussion until 100\% consensus was reached. The screening phase following the  inclusion and exclusion criteria yielded 911 inclusions and 517 exclusions.

Included repositories were then classified into construction strata. Two reviewers independently assigned each eligible repository to a stratum based on its documented software purpose and executable evidence. Stratum assignment was treated as a separate nominal coding task across seven categories, S01-S07 as presented in Table \ref{tab:2}. For example, a repository whose primary function was Industry Foundation Classes (IFC) model validation was classified as S01, while one whose primary function was personal protective equipment (PPE) detection from site imagery was classified as S06. Inter-rater agreement for stratum assignment was 91\% (Cohen’s $\kappa$ = 0.88), and disagreements were resolved through discussion until consensus was reached. The full codebook is provided in Appendix A (Table A1).

\begin{table*}[!t]
\caption{Construction Strata Assignment}
\label{tab:2}
\centering
\scriptsize
\setlength{\tabcolsep}{3pt}
\renewcommand{\arraystretch}{1.12}
\begin{tabularx}{\textwidth}{@{}>{\raggedright\arraybackslash}X>{\raggedright\arraybackslash}X>{\raggedright\arraybackslash}X>{\raggedright\arraybackslash}X@{}}
\toprule
\textbf{Code} & \textbf{Construction function} & \textbf{Included software} & \textbf{Example classification cue} \\
\midrule
S01 & BIM, openBIM, and model checking & Building information modeling, model exchange, viewing, conversion, and validation & A repository whose main function is Industry Foundation Classes (IFC) model validation \\
S02 & Construction project management & Planning, estimating, scheduling, costing, project documentation, and field management & A tool primarily used to schedule construction work \\
S03 & Construction digital twins & Digital representations connected to changing physical, sensor, or project data & A system that updates a project model using live site data \\
S04 & Reality capture and site sensing & Photogrammetry, LiDAR, point clouds, and site sensing & Software primarily used to process site imagery into a point cloud \\
S05 & Construction robotics and automation & Robotics, autonomous equipment, and automated fabrication & Software that controls construction equipment or automates fabrication \\
S06 & Jobsite IoT, monitoring, and safety & Connected site devices, monitoring, worker safety, and personal protective equipment analytics & A system whose main function is detecting personal protective equipment from site imagery \\
S07 & Multiple construction functions & Two or more of the functions above are central to the repository & A platform that integrates BIM model checking with project scheduling \\
\bottomrule
\end{tabularx}
\end{table*}

\subsection{Duplicate Adjudication and Repository Snapshotting}

To prevent redundant data from biasing the analysis, potential repository duplicates were formally evaluated and filtered using content-level comparison prior to assessing security outcomes. A repository was confirmed as a duplicate when its root-tree Git SHA matched that of another included repository. A pair was classified as a probable near-duplicate when it shared at least five blobs with containment of at least 0.98, or at least ten blobs with a Jaccard similarity of at least 0.90. Additionally, if repositories were explicitly identified as sequential stages of a single project, they were consolidated into a single project lineage based on commit chronology, provenance, and completeness. Seven repositories were removed under these rules, leaving the 904 repositories for further analysis with one Git tree snapshot per repository.

Next, for each of the 904 repositories, the default branch was resolved to an exact commit SHA. Its complete recursive Git tree was retrieved through the GitHub Git Trees API \cite{ref17}, with none of the tree responses truncated. Raw API responses, request metadata, input files, and SHA-256 hashes were retained for all repositories.

\subsection{Dependency Input Selection and Scanning}

Across the 904 frozen trees, 5,603 file paths matched the dependency manifest taxonomy. Dependency inputs were identified in the frozen trees through a file taxonomy (Appendix B, Table B1) that separated lockfiles and manifests from generated output, vendored dependency trees, cache directories, and virtual-environment files, the latter were discarded before analysis. A deterministic priority rule then selected one or more informative inputs per supported ecosystem from the 5,603 matching paths, preferring committed lockfiles when available and retaining manifests when they were the only supported evidence. For example, when a JavaScript repository contained both package.json and package-lock.json, the lockfile, which supplies resolved versions, was used for the primary scan. The rule selected 1,223 inputs across 529 repositories, and the corresponding Git blobs were downloaded and hash-verified. Original file names were preserved so that scanner parsers could recognize each input.

Selected inputs were scanned with OSV-Scanner version 2.5.1 \cite{ref20} in recursive JSON mode with all supported ecosystems enabled. We used OSV-Scanner because it's widely used, open-source, and reproducible. It matches exact package versions against ecosystem-native advisories in the OSV database across all ecosystems in our cohort, and it reports advisory metadata (i.e., severity, CWE). After scanning, out of the 529 repositories with selected inputs, 494 produced parseable results. The remaining 35 comprised 30 no-package cases and five scanner-error cases (details are reported in Table 4). The provenance record is provided in Appendix D. The 494 completed scans form the primary analytic population.

Freezing dependency inputs at a fixed commit ensured file-level reproducibility but did not necessarily preserve the resolved dependency graph across all ecosystems. Committed lockfiles generally record specific dependency versions, whereas manifests specify direct dependencies or version ranges. OSV-Scanner v2.5.1 supports transitive dependency resolution for Maven pom.xml and Python requirements.txt through external dependency metadata services. However, successful resolution depends on input format, configuration, and availability of external metadata. Consequently, dependencies resolved during scanning may reflect the state of external package metadata at the time of collection rather than the original repository commit. To assess differences in dependency identification and measurement coverage, lockfile-based results are reported separately as a sensitivity analysis \cite{ref20}.

\subsection{Constructs, Outcomes, and Analytic Populations}

Table \ref{tab:3} presents the constructs and populations used throughout the paper. The repository counts that populate these definitions are reported in Table \ref{tab:4} and Figure \ref{fig:flow}. Assessability was operationalized at two levels: input availability (repository contained at least one supported dependency input), and scan completion (primary scan succeeded and parsed at least one package).

\begin{table*}[!t]
\caption{Construct and Population Definitions.}
\label{tab:3}
\centering
\scriptsize
\setlength{\tabcolsep}{3pt}
\renewcommand{\arraystretch}{1.12}
\begin{tabularx}{\textwidth}{@{}>{\raggedright\arraybackslash}X>{\raggedright\arraybackslash}X>{\raggedright\arraybackslash}X@{}}
\toprule
\textbf{Term} & \textbf{Operational definition} & \textbf{Counts} \\
\midrule
Candidate & Unique repository returned by at least one of the 26 queries & 1,428 \\
Included & Candidate that passed eligibility screening & 911 \\
Analytic cohort & Included repository retained after duplicate adjudication and frozen at a commit SHA & 904 \\
Input available & Cohort repository with at least one supported dependency input selected for scanning (assessability, level 1) & 529 \\
Scan completed & Input-available repository whose primary scan succeeded and parsed at least one package (assessability, level 2; primary analytic population) & 494 \\
Affected & Scan-completed repository with at least one dependency version matched to a known vulnerability advisory & 395 \\
Advisory cluster & Connected set of OSV records linked by shared identifiers or aliases, counted once & 3,236 \\
Package family & Ecosystem-qualified package name, counted across versions & Affected repositories \\
\bottomrule
\end{tabularx}
\end{table*}

The primary outcome was a binary repository-level indicator of at least one dependency version matched to a known vulnerability advisory. The primary analytic population comprised the 494 completed scans. Repositories without a supported input, with no parsed packages, or with a scanner error were not treated as unaffected, because the absence of a finding in software that could not be observed is not evidence of the absence of exposure. A secondary sensitivity denominator counted the 30 no-package repositories as unaffected, while the full 904-repository fraction is reported only as an observed lower bound.

\subsection{Vulnerability Clustering and Severity}

Scanner occurrences were deduplicated by repository, path, ecosystem, package, version, and advisory. Because one vulnerability can appear under several identifiers, such as a CVE and a GitHub Security Advisory, records were merged into connected components whenever OSV identifiers or aliases overlapped. Each component received a canonical label that prioritized CVE, GHSA, PYSEC, RUSTSEC, GO, and OSV identifiers. From 36,379 raw scanner group occurrences and 191,621 parsed package rows, this procedure yielded 35,836 deduplicated occurrences and 3,236 unique advisory clusters.

Cluster severity was derived from OSV-Scanner’s numeric max\_severity values in the scanner output. For each repository-cluster occurrence, the scanner’s maximum numeric severity was retained. After alias-linked records consolidating, cluster-level severity was defined as the maximum usable numeric score across the cluster’s occurrences. Numeric values were mapped to the standard CVSS qualitative thresholds: \textit{Critical} (9.0-10.0), \textit{High} (7.0-8.9), \textit{\textit{Medium}} (4.0-6.9), and \textit{Low (0.1-3.9)} \cite{ref15} [Appendix C, Table C1]. Clusters lacking a usable numeric max\_severity were coded as \textit{Unknown}. Each affected repository was assigned the highest severity band among its advisory clusters. A repository was classified as \textit{Unknown} when none of its matched clusters had a usable numeric severity score. Advisories were not rescored, and the bands are 0.0 to 10.0 that measures the technical severity of a security vulnerability in software.

\subsection{Common Weakness Enumeration (CWE) Mapping}

Following advisory clustering and severity analysis, we mapped vulnerabilities to Common Weakness Enumeration (CWE) categories to characterize their underlying weakness types. We extracted explicit database\_specific.cwe\_ids from the retained OSV advisory records and associated them with alias-merged vulnerability clusters using canonical identifiers and advisory aliases \cite{ref30}. Each cluster was assigned the union of its explicitly reported CWE identifiers, without inferring additional labels from package names, CVSS scores, or advisory descriptions. Advisories lacking explicit CWE identifiers remained unmapped. We retrieved the corresponding weakness names from the MITRE CWE catalog \cite{ref28}.

The alias-merged vulnerability cluster served as the primary unit of CWE analysis, while affected-repository reach was used to assess prevalence. This approach avoids counting multiple advisory identifiers for the same vulnerability as distinct weaknesses. CWE categories were treated as nonexclusive labels describing the weakness types reported in advisory databases, whereas CVSS was analyzed separately as a measure of technical severity. Thus, the mapping characterizes documented weaknesses in reusable software components and classify weaknesses in reusable components.

\subsection{Vulnerability Burden and Shared Exposure}

Vulnerability burden among affected repositories was summarized with four counting units: unique advisory clusters, affected package families, affected package versions, and deduplicated occurrences. For each unit, we report the mean, standard deviation, median, interquartile range, 90th percentile, maximum, and Gini coefficient \cite{ref11}. These units answer different questions. Clusters describe distinct advisory entities, package families identify shared remediation targets, versions describe affected releases, and deduplicated occurrences describe repository-level exposure without counting repeated aliases as independent vulnerabilities.

\subsection{Repository Characteristics and Covariates}

To examine whether observable repository management practices are associated with vulnerable dependency exposure, we analyzed six configuration-based artifacts identified from the collected Git tree inventories: dependency-update automation (e.g., Dependabot or Renovate), continuous integration (CI), security policy files, container build files, CODEOWNERS files, and pre-commit configurations. These artifacts capture different aspects of repository infrastructure, including dependency management, development automation, security governance, ownership configuration, and build processes. We selected these indicators because they can be identified directly from repository snapshots using reproducible file-based criteria, enabling consistent comparisons across repositories without relying on historical development activity.

For each repository, we recorded the presence or absence of each artifact and constructed a composite binary indicator representing the presence of at least one of the six. The individual indicators allowed us to examine artifact-specific associations, while the composite captured the broader presence of repository management and automation infrastructure. Because CI configurations and container build files accounted for a substantial proportion of composite positives, we interpreted the composite as a general configuration indicator rather than a measure of security maturity or maintenance effectiveness.

To account for observable repository-level differences, we considered repository age, inactivity (time since the last push), size, star count, owner type (individual or organization), archive status, and construction stratum as covariates. For scanned repositories, we additionally considered parsed package count to account for differences in the number of identified dependencies. We examined associations between artifact presence and the likelihood of detecting at least one known vulnerability advisory match, reporting odds ratios (ORs), 95\% confidence intervals (CIs), and multiple-testing-adjusted q-values. These analyses characterize associations between observable repository configurations and detected vulnerability exposure, rather than causal effects of maintenance practices.

\subsection{Statistical Analysis}

We probed the sensitivity of our central estimate through several statistical analyses. This analyses recomputed the primary proportion after counting no-package repositories as unaffected, after restricting to production-proxy paths, which excluded dependency files whose paths contained test, example, sample, demo, fixture, documentation, or benchmark markers, after restrictions by archive status, recent activity, stars, and owner type and separately for lockfile inputs. The five scanner-error repairs were retained as a diagnostic reproducibility record in Appendix D and were excluded from the primary and reported robustness estimates. All proportions are reported with two-sided 95\% Wilson score confidence intervals. Strata with fewer than 15 scanned repositories are interpreted descriptively. Because the design is cross-sectional, all associations are treated as descriptive or exploratory rather than causal.

To characterize differential assessability, we fitted logistic regressions for input availability and scan completion on 904 cohort repositories. Predictors were standardized repository age and inactivity (time since the last push), log(1 + size), log(1 + stars), owner type, archive status, and stratum with S01 as the reference, and standard errors were heteroskedasticity-consistent \cite{ref48}. These models describe which repositories are observable, they were not used to impute missing security outcomes.

Prevalence and assessability differences across strata in input availability and in the proportion of scanned repositories affected were assessed with Pearson chi-square tests, with Cramér’s V as the effect size. Since several expected cell counts were below five, p-values were also obtained by Monte Carlo simulation of the exact conditional test with fixed margins. Ecosystem results are reported per ecosystem; because one repository can contain several ecosystems, ecosystem rows overlap.

A greedy set-cover algorithm ranked package families by the number of not-yet-covered affected repositories each would add, using a standard approximation for the set-cover problem \cite{ref10,ref24}. Ties were broken by the total number of affected repositories containing the family, in descending order. The ranking maximizes repository reach while it does not account for severity, reachability, or remediation feasibility.

Associations between each artifact and the primary outcome were estimated among the 494 completed scans using separate Firth bias-reduced logistic regressions, which reduce small-sample bias and remain estimable under separation \cite{ref16,ref21}. Each model adjusted for log(1 + package count), owner type, log(1 + stars), and construction stratum. As several strata had sparse completed-scan counts, strata were represented using S01, S02, and S06 as separate categories and an ``Other strata'' category combining S03, S04, S05, and S07. This pooling was used only for regression adjustment, all descriptive and stratified results retained the original seven-stratum classification. Odds ratios used Wald-type 95\% CIs, and p-values were adjusted across the seven artifact tests with the Benjamini-Hochberg false-discovery-rate procedure \cite{ref6}. Given the sparseness of the data, these cross-sectional associations are treated as exploratory rather than causal.

\section{Results}

\subsection{Dataset Construction and Cohort Characteristics}

\begin{figure}[!t]
\centering
\includegraphics[width=\columnwidth]{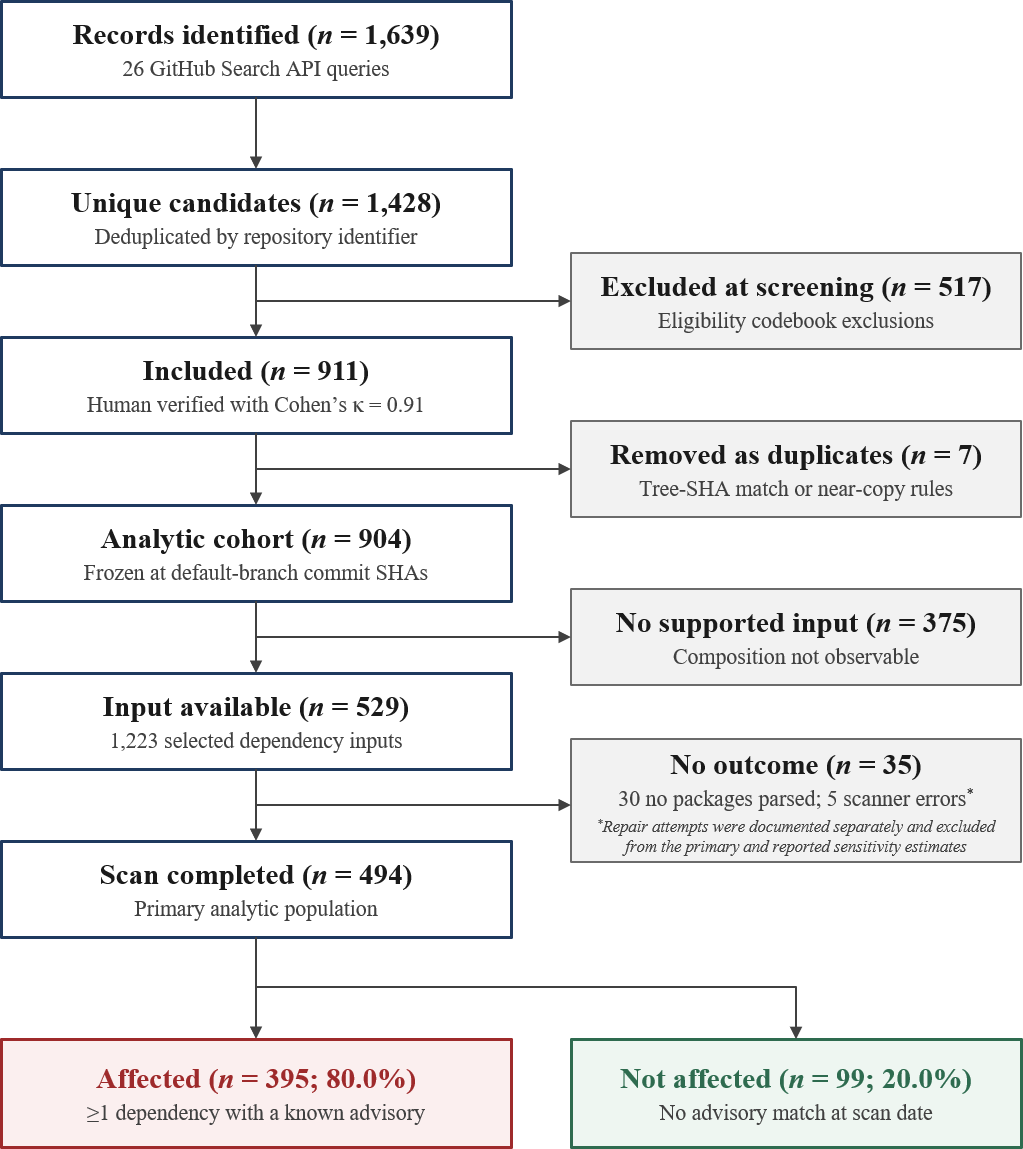}
\caption{Repository identification, screening, duplicate adjudication, and dependency-scan flow}
\label{fig:flow}
\end{figure}

The 26 search queries yielded 1,639 records corresponding to 1,428 unique repositories (Figure \ref{fig:flow}). Following eligibility screening, 911 repositories were retained and 517 were excluded. Subsequent duplicate adjudication removed seven additional repositories, resulting in a final cohort of 904 repositories. Each repository was anchored to a specific commit SHA, with its complete Git tree inventory retrieved to establish a reproducible snapshot for analysis.

The resulting cohort was concentrated in three construction subdomains: construction project management (S02; 280 repositories, 31.0\%), BIM/openBIM/model checking (S01; 248, 27.4\%), and jobsite IoT/monitoring/safety (S06; 206, 22.8\%). Together, these categories accounted for 81.2\% of the cohort. The remaining repositories covered robotics/automation (S05; 70, 7.7\%), multiple construction functions (S07; 65, 7.2\%), digital twins (S03; 27, 3.0\%), and reality capture/site sensing (S04; 8, 0.9\%).

Overall, the dataset spans seven construction-related subdomains but is heavily represented by project management, BIM, and jobsite monitoring applications, a distribution that should be considered when interpreting aggregate findings across the cohort.

\subsection{RQ1: Repository Assessability}

Of the 904 repositories in the cohort, 529 (58.5\%, 95\% CI: 55.3–61.7\%) contained at least one supported dependency input, and 494 (54.6\%, 95\% CI: 51.4–57.9\%) completed the primary vulnerability scan (Table 4). The remaining 410 repositories could not be assessed through the primary scanning pipeline: 375 lacked supported dependency inputs, 30 had inputs that yielded no parsed packages, and five encountered scanner errors. Consequently, the 494 successfully scanned repositories constitute the analytic population for RQ2–RQ5, representing just over half of the original cohort.

\begin{table*}[!t]
\caption{Cohort and dependency-scan disposition.}
\label{tab:4}
\centering
\scriptsize
\setlength{\tabcolsep}{3pt}
\renewcommand{\arraystretch}{1.12}
\begin{tabularx}{\textwidth}{@{}>{\raggedright\arraybackslash}X>{\raggedright\arraybackslash}X>{\raggedright\arraybackslash}X@{}}
\toprule
\textbf{Disposition} & \textbf{Repositories} & \textbf{\% of cohort} \\
\midrule
Analytic cohort & 904 & 100.0 \\
Input available (at least one supported dependency input) & 529 & 58.5 \\
No supported dependency input & 375 & 41.5 \\
Scan completed (primary analytic population) & 494 & 54.6 \\
Affected (at least one known-vulnerable dependency) & 395 & 43.7 \\
No advisory match detected & 99 & 11.0 \\
No packages parsed & 30 & 3.3 \\
Scanner error (retained as a sensitivity record) & 5 & 0.6 \\
\bottomrule
\end{tabularx}
\vspace{4pt}
\begin{minipage}{\textwidth}\scriptsize Percentages use the 904-repository cohort. The affected share of the cohort (43.7\%) is an observed minimum, not a prevalence estimate, because 410 repositories had no outcome.\end{minipage}
\end{table*}
Dependency assessability varied significantly across construction subdomains ($\chi^2$ = 54.96, df = 6, Monte Carlo p < 0.001, Cramér’s V = 0.247; Table 5, Figure. 2). Supported dependency inputs were most prevalent in construction project management (67.9\%) and BIM/model checking (62.5\%). In contrast, only 15 of the 70 robotics and automation repositories (21.4\%) contained a supported input. This disparity may reflect differences in the dependency management ecosystems used across construction applications, as well as the scanner's supported input formats. In particular, robotics workflows commonly rely on ROS package manifests, C/C++ build systems, Conan, vcpkg, conda, or Unity package files that were not supported as first-class dependency inputs by the scanning configuration used in this study \cite{ref20}

In the scan-completion model, a one-standard-deviation increase in inactivity was associated with lower odds of a completed scan (odds ratio [OR] = 0.63, 95\% CI 0.49-0.82, p < 0.001), whereas larger repository size (OR = 1.26, 95\% CI 1.07-1.48, p = 0.006) and organizational ownership (OR = 1.77, 95\% CI 1.09-2.86, p = 0.020) were associated with higher odds. Archived repositories also had higher modeled odds (OR = 5.18, 95\% CI 1.89-14.18, p = 0.001), although only 22 cohort repositories were archived. Relative to S01, completion odds were higher in S02 (OR = 1.73, 95\% CI 1.10-2.74) and lower in S05 (OR = 0.20, 95\% CI 0.10-0.39) and S07 (OR = 0.50, 95\% CI 0.27-0.93). Completed scans were therefore not a random cross-ecosystem  of the cohort. Details of finding for each stratum is provided in Table \ref{tab:5}.

\begin{table*}[!t]
\caption{Input availability and detected vulnerability by construction stratum.}
\label{tab:5}
\centering
\scriptsize
\setlength{\tabcolsep}{3pt}
\renewcommand{\arraystretch}{1.12}
\begin{tabularx}{\textwidth}{@{}>{\raggedright\arraybackslash}X>{\raggedright\arraybackslash}X>{\raggedright\arraybackslash}X>{\raggedright\arraybackslash}X>{\raggedright\arraybackslash}X>{\raggedright\arraybackslash}X>{\raggedright\arraybackslash}X@{}}
\toprule
\textbf{Stratum} & \textbf{Cohort (n)} & \textbf{Input available (n)} & \textbf{Input available \% (95\% CI)} & \textbf{Scanned (n)} & \textbf{Affected (n)} & \textbf{Affected \% (95\% CI)} \\
\midrule
S01 & 248 & 155 & 62.5 (56.3-68.3) & 139 & 101 & 72.7 (64.7-79.4) \\
S02 & 280 & 190 & 67.9 (62.2-73.1) & 183 & 160 & 87.4 (81.8-91.5) \\
S03 & 27 & 14 & 51.9 (34.0-69.3) & 12 & 11 & 91.7 (64.6-98.5) \\
S04 & 8 & 3 & 37.5 (13.7-69.4) & 2 & 0 & 0.0 (0.0-65.8) \\
S05 & 70 & 15 & 21.4 (13.4-32.4) & 13 & 7 & 53.8 (29.1-76.8) \\
S06 & 206 & 119 & 57.8 (50.9-64.3) & 114 & 88 & 77.2 (68.7-83.9) \\
S07 & 65 & 33 & 50.8 (38.9-62.5) & 31 & 28 & 90.3 (75.1-96.7) \\
All strata & 904 & 529 & 58.5 (55.3-61.7) & 494 & 395 & 80.0 (76.2-83.3) \\
\bottomrule
\end{tabularx}
\vspace{4pt}
\begin{minipage}{\textwidth}\scriptsize Input-availability denominators are stratum cohorts; affected denominators are completed scans within each stratum. Wilson 95\% CIs. Input availability: $\chi^2$ = 54.96, df = 6, V = 0.247. Affected: $\chi^2$ = 28.16, df = 6, V = 0.239. Both Monte Carlo p < 0.001 (100,000 replicates). S03-S05 had fewer than 15 scanned repositories and are interpreted descriptively.\end{minipage}
\end{table*}

\begin{figure*}[!t]
\centering
\includegraphics[width=0.98\textwidth]{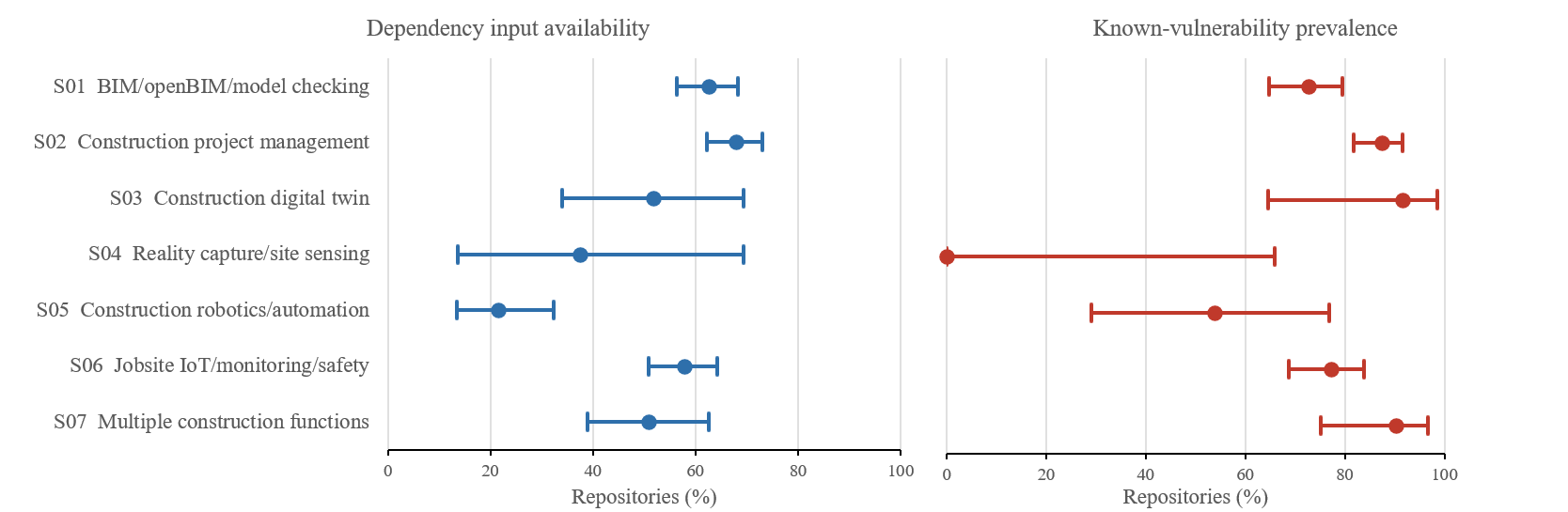}
\caption{Input availability (left; denominator = stratum cohort) and known-vulnerable prevalence (right; denominator = completed scans) by construction stratum. Points are proportions; bars are 95\% Wilson CI.}
\label{fig:strata}
\end{figure*}

\subsection{RQ2: Prevalence of Known-Vulnerable Dependencies}
Known-vulnerable dependencies were detected in most successfully scanned repositories. Of the 494 repositories that completed the primary scan, 395 (80.0\%, 95\% CI: 76.2–83.3\%) contained at least one dependency version matched to a published vulnerability advisory (Table 6). This estimate remained nearly unchanged when restricting the analysis to production-proxy paths (393/492, 79.9\%, 95\% CI: 76.1–83.2\%). Including the 30 repositories with supported inputs but no parsed packages as unaffected reduced the observed proportion to 75.4\% (395/524, 95\% CI: 71.5–78.9\%). Across the full 904-repository cohort, confirmed advisory matches were observed in at least 43.7\% (395/904, 95\% CI: 40.5–46.9\%), representing a lower bound on observed exposure rather than an estimate of cohort-wide vulnerability prevalence.

Observed prevalence was sensitive to dependency input type. Restricting the analysis to repositories with lockfile inputs produced the highest observed prevalence among the reported sensitivity analyses, with 253 of 263 repositories (96.2\%, 95\% CI: 93.1–97.9\%) containing at least one advisory match. This proportion was 16.2 percentage points higher than the primary estimate. The difference reflects more complete dependency information in lockfiles, as well as differences in repository and ecosystem composition. Therefore, the analysis does not establish that input type alone accounts for the observed variation.

\begin{table*}[!t]
\caption{Primary estimate and selected robustness analyses.}
\label{tab:6}
\centering
\scriptsize
\setlength{\tabcolsep}{3pt}
\renewcommand{\arraystretch}{1.12}
\begin{tabularx}{\textwidth}{@{}>{\raggedright\arraybackslash}X>{\raggedright\arraybackslash}X>{\raggedright\arraybackslash}X>{\raggedright\arraybackslash}X>{\raggedright\arraybackslash}X@{}}
\toprule
\textbf{Analysis} & \textbf{n} & \textbf{Affected} & \textbf{\% affected (95\% CI)} & \textbf{$\Delta$ from primary (pp)} \\
\midrule
Primary: repositories with a completed scan & 494 & 395 & 80.0 (76.2-83.3) & 0.0 \\
Counting no-package repositories as not affected & 524 & 395 & 75.4 (71.5-78.9) & \textminus{}4.6 \\
Production-proxy dependency paths & 492 & 393 & 79.9 (76.1-83.2) & \textminus{}0.1 \\
Non-archived repositories & 477 & 379 & 79.5 (75.6-82.8) & \textminus{}0.5 \\
Pushed within five years & 474 & 378 & 79.7 (75.9-83.1) & \textminus{}0.2 \\
Frozen commit within five years & 463 & 367 & 79.3 (75.3-82.7) & \textminus{}0.7 \\
At least one star & 239 & 189 & 79.1 (73.5-83.8) & \textminus{}0.9 \\
Organization-owned & 82 & 66 & 80.5 (70.6-87.6) & +0.5 \\
User-owned & 412 & 329 & 79.9 (75.7-83.4) & \textminus{}0.1 \\
Lockfile inputs only & 263 & 253 & 96.2 (93.1-97.9) & +16.2 \\
\bottomrule
\end{tabularx}
\vspace{4pt}
\begin{minipage}{\textwidth}\scriptsize pp = percentage points; Wilson 95\% CIs. Production-proxy paths exclude dependency files whose paths contain test, example, sample, demo, fixture, documentation, or benchmark markers. The lockfile-only estimate conditions on repositories with supported lockfile inputs and is not directly comparable to the primary population, because lockfiles enumerate resolved transitive dependencies.\end{minipage}
\end{table*}

Vulnerability exposure varied significantly across construction subdomains. Among successfully scanned repositories, the proportion with at least one advisory match differed by construction stratum ($\chi^2$ = 28.16, df = 6, Monte Carlo p < 0.001, V = 0.239), as shown in Table \ref{tab:5}, Figure \ref{fig:strata}. The highest observed proportions occurred in repositories serving multiple construction functions (90.3\%) and construction project management (87.4\%), compared with BIM/model checking (72.7\%) and robotics/automation (53.8\%). Neither of the two scanned reality-capture repositories had a detected advisory match. Excluding this sparsely represented stratum, the association remained statistically significant ($\chi^2$ = 20.39, df = 5, Monte Carlo p = 0.002, V = 0.204). Given the small denominators in S03-S05, stratum differences in prevalence are descriptive.

Detected exposure also differed across package ecosystems. As shown in Table \ref{tab:7}, npm had the highest observed prevalence among the more frequently represented ecosystems, with 231 of 239 repositories (96.7\%) containing at least one advisory match. Corresponding proportions were 71.4\% for PyPI (150/210), 29.9\% for NuGet (20/67), and 74.2\% for Maven (23/31). Packagist also exhibited a high observed proportion (17/18, 94.4\%). These differences are consistent with variation in dependency representation and ecosystem composition.

Overall, known-vulnerable dependencies were detected in 80.0\% of successfully scanned construction-related repositories, indicating substantial observable exposure within the assessable subset. However, prevalence varied across dependency input types, construction subdomains, and package ecosystems. These findings demonstrate that measured vulnerability exposure depends not only on the dependencies present but also on which repositories and dependency relationships can be assessed. Consequently, the estimates characterize detectable known-vulnerability exposure rather than the underlying vulnerability prevalence of the entire construction-software cohort.

\begin{table*}[!t]
\caption{Package ecosystems in completed scans.}
\label{tab:7}
\centering
\scriptsize
\setlength{\tabcolsep}{3pt}
\renewcommand{\arraystretch}{1.12}
\begin{tabularx}{\textwidth}{@{}>{\raggedright\arraybackslash}X>{\raggedright\arraybackslash}X>{\raggedright\arraybackslash}X>{\raggedright\arraybackslash}X>{\raggedright\arraybackslash}X@{}}
\toprule
\textbf{Ecosystem} & \textbf{Repositories with parsed packages} & \textbf{Affected repositories} & \textbf{Affected \% (95\% Wilson CI)} & \textbf{Unique advisory clusters} \\
\midrule
npm & 239 & 231 & 96.7 (93.5-98.3) & 1,099 \\
PyPI & 210 & 150 & 71.4 (65.0-77.1) & 1,254 \\
NuGet & 67 & 20 & 29.9 (20.2-41.7) & 85 \\
Maven & 31 & 23 & 74.2 (56.8-86.3) & 414 \\
Packagist & 18 & 17 & 94.4 (74.2-99.0) & 193 \\
crates.io & 11 & 9 & 81.8 (52.3-94.9) & 63 \\
Go & 3 & 3 & 100.0 (43.9-100.0) & 70 \\
Pub & 2 & 0 & 0.0 (0.0-65.8) & 0 \\
RubyGems & 2 & 2 & 100.0 (34.2-100.0) & 66 \\
SwiftURL & 1 & 0 & 0.0 (0.0-79.3) & 0 \\
\bottomrule
\end{tabularx}
\vspace{4pt}
\begin{minipage}{\textwidth}\scriptsize Rows are not mutually exclusive, because multi-ecosystem repositories contribute to several rows. Cluster counts sum to 3,244 rather than 3,236 because some advisory clusters span more than one ecosystem.\end{minipage}
\end{table*}

\subsection{RQ3: Vulnerability Severity and Security Weakness}

\textit{High} and \textit{Critical}-severity vulnerabilities accounted for nearly half of the unique advisory clusters. Among the 3,236 alias-merged vulnerability clusters, 1,276 (39.4\%) were classified as \textit{High} severity and 318 (9.8\%) as \textit{Critical}, collectively representing 49.3\% of the clusters. \textit{Medium} severity vulnerabilities accounted for 1,207 clusters (37.3\%), followed by Low (265, 8.2\%) and \textit{Unknown} (170, 5.3\%). Across 28,085 unique repository–cluster pairs, \textit{High} severity vulnerabilities were the most frequently observed (13,317 pairs, 47.4\%), followed by \textit{Medium} severity vulnerabilities (10,766, 38.3\%), indicating that these severity categories accounted for most observed repository–vulnerability associations.

\begin{figure}[!t]
\centering
\includegraphics[width=\columnwidth]{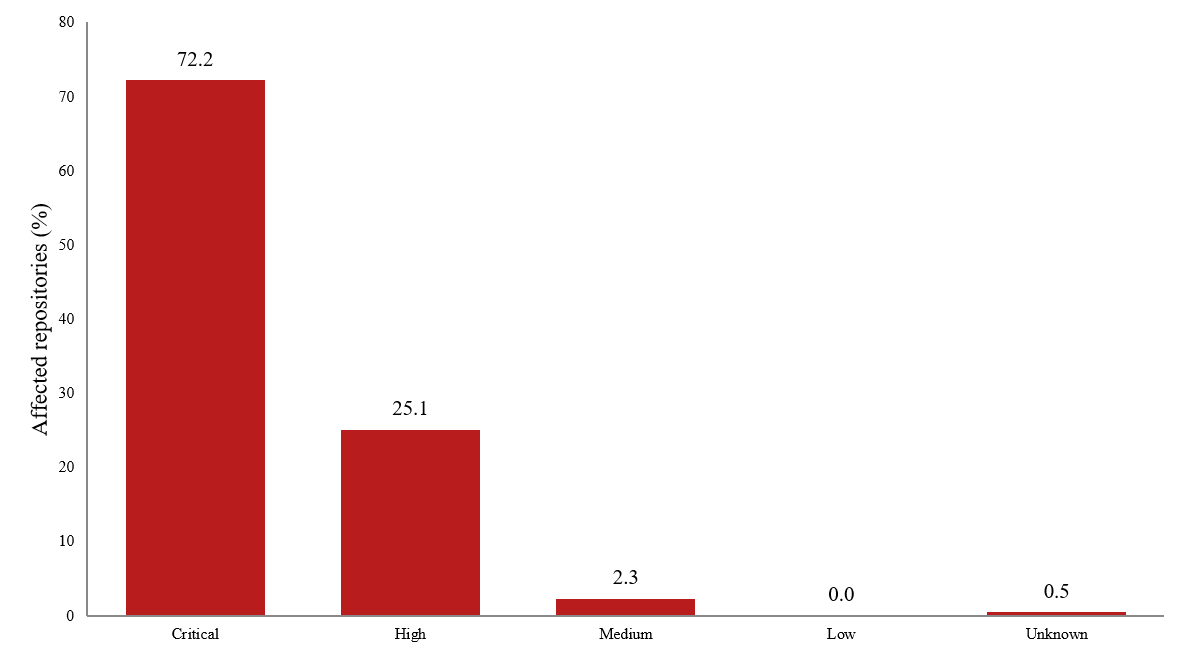}
\caption{Maximum advisory severity among the 395 affected repositories.}
\label{fig:severity}
\end{figure}

Most affected repositories contained at least one \textit{High} or \textit{Critical}-severity vulnerability. When repositories were classified by their maximum detected severity, 285 of the 395 affected repositories (72.2\%, 95\% CI: 67.5–76.3\%) had at least one \textit{Critical}-severity match, while 99 (25.1\%) had \textit{High} as their maximum severity. Only nine repositories (2.3\%) had \textit{Medium} as their maximum severity, none had \textit{Low}, and two (0.5\%) had \textit{Unknown} severity. Overall, 384 affected repositories (97.2\%, 95\% CI: 95.1–98.4\%) contained at least one vulnerability classified as \textit{High} or \textit{Critical} (Table \ref{tab:8}, Figure. \ref{fig:severity}). Thus, although \textit{Critical} vulnerabilities represented a relatively small fraction of distinct clusters, they were distributed across a substantial proportion of affected repositories.

\begin{table*}[!t]
\caption{Advisory severity at two counting levels.}
\label{tab:8}
\centering
\scriptsize
\setlength{\tabcolsep}{3pt}
\renewcommand{\arraystretch}{1.12}
\begin{tabularx}{\textwidth}{@{}>{\raggedright\arraybackslash}X>{\raggedright\arraybackslash}X>{\raggedright\arraybackslash}X>{\raggedright\arraybackslash}X>{\raggedright\arraybackslash}X>{\raggedright\arraybackslash}X>{\raggedright\arraybackslash}X@{}}
\toprule
\textbf{Counting level} & \textbf{Critical} & \textbf{High} & \textbf{Medium} & \textbf{Low} & \textbf{Unknown} & \textbf{Total} \\
\midrule
Advisory clusters & 318 (9.8\%) & 1,276 (39.4\%) & 1,207 (37.3\%) & 265 (8.2\%) & 170 (5.3\%) & 3,236 \\
Affected-repository maximum severity & 285 (72.2\%) & 99 (25.1\%) & 9 (2.3\%) & 0 (0.0\%) & 2 (0.5\%) & 395 \\
\bottomrule
\end{tabularx}
\vspace{4pt}
\begin{minipage}{\textwidth}\scriptsize At the repository-cluster pair level (28,085 unique pairs), High was the most common band (13,317 pairs, 47.4\%), followed by \textit{Medium} (10,766 pairs, 38.3\%). Severity bands follow the CVSS qualitative thresholds (Critical 9.0-10.0, High 7.0-8.9, \textit{Medium} 4.0-6.9, \textit{Low} 0.1-3.9); ``\textit{Unknown}'' = no usable numeric score.\end{minipage}
\end{table*}

Security weakness classification revealed widespread resource-consumption and input-handling weaknesses. Explicit CWE metadata was available in 35,224 of the 43,959 nested advisory records (80.1\%). After alias-aware linkage, 3,069 of the 3,236 vulnerability clusters (94.8\%) had at least one CWE identifier, while 167 clusters remained unmapped. In total, 291 distinct CWE identifiers were observed. The categories with the broadest affected-repository reach were CWE-770 (Allocation of Resources Without Limits or Throttling; 338 repositories, 85.6\%), CWE-400 (Uncontrolled Resource Consumption; 337, 85.3\%), CWE-835 (Loop with Unreachable Exit Condition, or Infinite Loop; 299, 75.7\%), and CWE-1333 (Inefficient Regular Expression Complexity; 292, 73.9\%). Other widely observed categories included CWE-22 (Path Traversal; 280, 70.9\%) and CWE-200 (Exposure of Sensitive Information to an Unauthorized Actor; 259, 65.6\%). Table \ref{tab:9} presents the ten categories with the greatest repository reach, while Figure. 4 illustrates their distribution across construction strata.
Code injection also represented a notable high-severity weakness category. CWE-94 (Improper Control of Generation of Code, or Code Injection) was associated with vulnerabilities affecting 234 repositories. Of the 83 vulnerability clusters mapped to this category, 70 (84.3\%) were classified as \textit{High} or \textit{Critical}, with a median numeric cluster severity of 8.80. CWE categories were not mutually exclusive, as a vulnerability cluster could contain multiple explicitly reported weakness identifiers. For example, CWE-400 and CWE-770 co-occurred in 50 clusters. Accordingly, repository reach reflects the presence of advisory-reported weakness categories.

Overall, \textit{High} and \textit{Critical} severity vulnerabilities accounted for 49.3\% of unique vulnerability clusters, yet 97.2\% of affected repositories contained at least one vulnerability in these severity bands. Resource-consumption weaknesses had the broadest repository reach, while code injection was also widespread and predominantly associated with high severity clusters. Together, these findings reveal substantial exposure to severe, known vulnerabilities and recurring weakness categories among the scanned construction-related repositories.

\begin{table*}[!t]
\caption{CWE types ranked by affected-repository reach.}
\label{tab:9}
\centering
\scriptsize
\setlength{\tabcolsep}{3pt}
\renewcommand{\arraystretch}{1.12}
\begin{tabularx}{\textwidth}{@{}>{\raggedright\arraybackslash}X>{\raggedright\arraybackslash}X>{\raggedright\arraybackslash}X>{\raggedright\arraybackslash}X>{\raggedright\arraybackslash}X>{\raggedright\arraybackslash}X@{}}
\toprule
\textbf{CWE} & \textbf{Short weakness label} & \textbf{Affected repositories} & \textbf{\% of affected repositories} & \textbf{Clusters} & \textbf{High or Critical clusters} \\
\midrule
CWE-770 & Resource allocation without limits & 338 & 85.6 & 155 & 88 \\
CWE-400 & Uncontrolled resource consumption & 337 & 85.3 & 297 & 177 \\
CWE-835 & Infinite loop & 299 & 75.7 & 49 & 27 \\
CWE-1333 & Inefficient regular-expression complexity & 292 & 73.9 & 123 & 67 \\
CWE-22 & Path traversal & 280 & 70.9 & 169 & 102 \\
CWE-200 & Sensitive-information exposure & 259 & 65.6 & 94 & 45 \\
CWE-190 & Integer overflow or wraparound & 256 & 64.8 & 49 & 28 \\
CWE-407 & Inefficient algorithmic complexity & 235 & 59.5 & 53 & 36 \\
CWE-94 & Code injection & 234 & 59.2 & 83 & 70 \\
CWE-20 & Improper input validation & 222 & 56.2 & 187 & 69 \\
\bottomrule
\end{tabularx}
\vspace{4pt}
\begin{minipage}{\textwidth}\scriptsize The denominator for repository reach is 395 affected repositories. CWE categories are nonexclusive, so percentages must not be summed. \textit{High} or \textit{Critical} counts are cluster counts, not repository counts. Full MITRE names and all 291 identifiers are provided in the Appendix.\end{minipage}
\end{table*}

\begin{figure*}[!t]
\centering
\includegraphics[width=0.92\textwidth]{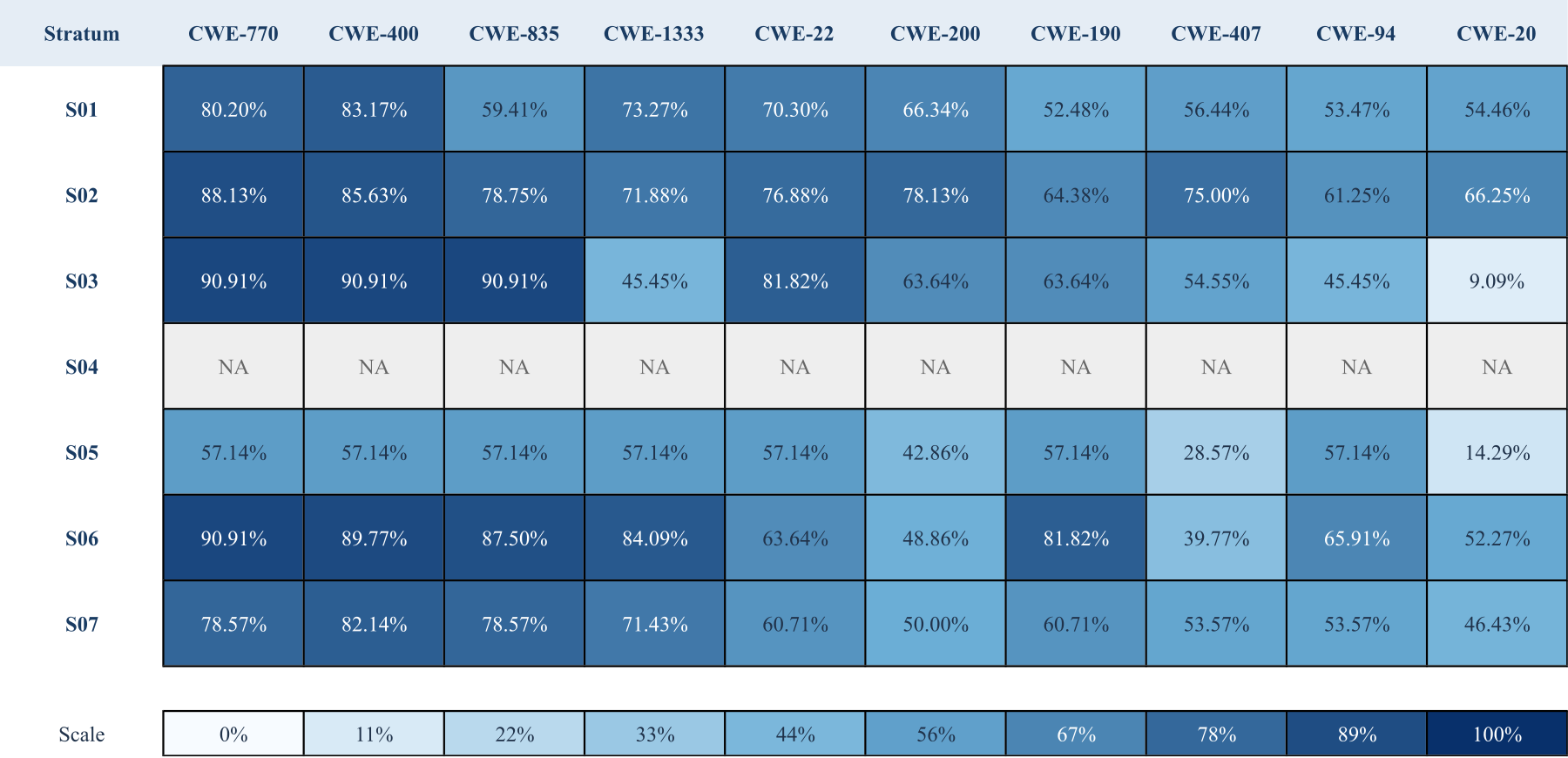}
\caption{Reach of the ten most common native CWE types across construction strata. S04 is marked NA because it contained no affected repositories; this is not a zero-risk estimate.}
\label{fig:cwe}
\end{figure*}

\subsection{RQ4: Burden and Shared Exposure}

Vulnerability burden was unevenly distributed across affected repositories. Among the 395 repositories with detected advisory matches, the median repository contained 50 distinct vulnerability clusters (IQR: 21–110; maximum: 507) spanning 12 affected package families (IQR: 5–28; maximum: 120). The distributions were strongly right-skewed, with Gini coefficients ranging from 0.492 for distinct vulnerability clusters to 0.578 for affected package versions (Table \ref{tab:10}). These results indicate substantial variation in the number of known-vulnerability matches across repositories, with a comparatively small subset accounting for a disproportionate share of the observed burden.

\begin{table*}[!t]
\caption{Vulnerability burden among the 395 affected repositories.}
\label{tab:10}
\centering
\scriptsize
\setlength{\tabcolsep}{3pt}
\renewcommand{\arraystretch}{1.12}
\begin{tabularx}{\textwidth}{@{}>{\raggedright\arraybackslash}X>{\raggedright\arraybackslash}X>{\raggedright\arraybackslash}X>{\raggedright\arraybackslash}X>{\raggedright\arraybackslash}X>{\raggedright\arraybackslash}X>{\raggedright\arraybackslash}X>{\raggedright\arraybackslash}X@{}}
\toprule
\textbf{Counting unit (per affected repository)} & \textbf{Mean} & \textbf{SD} & \textbf{Median} & \textbf{IQR} & \textbf{P90} & \textbf{Max} & \textbf{Gini} \\
\midrule
Unique advisory clusters & 71.1 & 67.6 & 50 & 21-110 & 160.8 & 507 & 0.492 \\
Affected package families & 20.0 & 22.6 & 12 & 5-28 & 49.0 & 120 & 0.544 \\
Affected package versions & 23.7 & 30.2 & 12 & 5-31 & 59.0 & 191 & 0.578 \\
Deduplicated occurrences & 90.7 & 109.8 & 54 & 23-120.5 & 204.0 & 725 & 0.548 \\
\bottomrule
\end{tabularx}
\vspace{4pt}
\begin{minipage}{\textwidth}\scriptsize Counting units are defined in Section IV.H. SD = standard deviation; P90 = 90th percentile. Higher Gini values indicate greater inequality of burden across affected repositories.\end{minipage}
\end{table*}

A small number of upstream package families accounted for widespread shared exposure. The most widely shared affected family, \texttt{npm}, appeared with at least one matched advisory in 183 of the 395 affected repositories (46.3\%). Including \texttt{PyPI} increased the cumulative reach to 273 repositories (69.1\%), representing 90 additional repositories not already covered by the first family. The five most widely shared affected families collectively reached 326 repositories (82.5\%), increasing to 357 (90.4\%) for the top ten and 376 (95.2\%) for the top twenty (Table 11). These cumulative counts represent distinct repositories affected by at least one of the selected package families, accounting for overlap between families.

Shared vulnerability exposure extended across multiple package ecosystems and software functions. The ten highest-ranked affected package families spanned npm, PyPI, Maven, NuGet, and Packagist, including libraries supporting stylesheet processing, domain-name and query-string handling, image processing, JSON serialization, web routing, and machine learning. This distribution indicates that shared exposure was not confined to a single package ecosystem or application function, but involved reusable components serving different roles within construction-related software.

Overall, the known-vulnerability burden was highly uneven across affected repositories, while exposure to a small number of shared upstream package families was widespread. Notably, the twenty most widely shared affected families collectively reached 95.2\% of affected repositories. These findings highlight the importance of distinguishing repository-specific vulnerability burden from shared dependency exposure, where vulnerabilities in widely reused components can be observed across many otherwise distinct repositories.

\begin{table*}[!t]
\caption{Ten highest-ranked package families under greedy repository coverage.}
\label{tab:11}
\centering
\scriptsize
\setlength{\tabcolsep}{3pt}
\renewcommand{\arraystretch}{1.12}
\begin{tabularx}{\textwidth}{@{}>{\raggedright\arraybackslash}X>{\raggedright\arraybackslash}X>{\raggedright\arraybackslash}X>{\raggedright\arraybackslash}X>{\raggedright\arraybackslash}X>{\raggedright\arraybackslash}X@{}}
\toprule
\textbf{Rank} & \textbf{Package family} & \textbf{Affected repositories} & \textbf{Newly covered} & \textbf{Cumulative (n)} & \textbf{Cumulative \%} \\
\midrule
1 & npm:\allowbreak postcss & 183 & 183 & 183 & 46.3 \\
2 & PyPI:\allowbreak idna & 110 & 90 & 273 & 69.1 \\
3 & npm:\allowbreak qs & 116 & 21 & 294 & 74.4 \\
4 & PyPI:\allowbreak pillow & 101 & 20 & 314 & 79.5 \\
5 & Maven:\allowbreak com.\allowbreak fasterxml.\allowbreak jackson.\allowbreak core:\allowbreak jackson-databind & 18 & 12 & 326 & 82.5 \\
6 & npm:\allowbreak brace-expansion & 173 & 10 & 336 & 85.1 \\
7 & NuGet:\allowbreak Newtonsoft.\allowbreak Json & 11 & 9 & 345 & 87.3 \\
8 & npm:\allowbreak nanoid & 175 & 4 & 349 & 88.4 \\
9 & PyPI:\allowbreak torch & 72 & 4 & 353 & 89.4 \\
10 & Packagist:\allowbreak symfony/\allowbreak routing & 15 & 4 & 357 & 90.4 \\
\bottomrule
\end{tabularx}
\vspace{4pt}
\begin{minipage}{\textwidth}\scriptsize Package families are ecosystem-qualified to prevent cross-ecosystem name collisions. Newly covered counts exclude repositories reached by higher-ranked families. Ranks 8-10 tie on newly covered repositories and are ordered by total affected repositories.\end{minipage}
\end{table*}

\subsection{RQ5: Repository Characteristics and Vulnerability Exposure}

Repository-level artifacts were present in a minority of the studied repositories. Among the 904 repositories in the cohort, 215 (23.8\%) contained at least one of the six examined artifacts. Of the 494 repositories included in the vulnerability analysis, 164 (33.2\%) contained at least one artifact. Continuous-integration (CI) configurations and container build files were the most prevalent, appearing in 103 and 77 scanned repositories, respectively. The combined presence of repository artifacts showed no statistically significant association with vulnerable dependency exposure. Repositories containing at least one of the examined artifacts had lower estimated odds of a detected vulnerability advisory match; however, the association did not reach statistical significance after multiple-testing correction (OR = 0.70, 95\% CI: 0.31–1.59, q = 0.461; Table 12). The confidence interval encompassed both lower and higher odds, indicating uncertainty regarding the direction and magnitude of the association.

Dependency-update automation was the only individual artifact significantly associated with lower vulnerability exposure. Repositories containing dependency-update automation configurations exhibited substantially lower estimated odds of a detected advisory match (OR = 0.037, 95\% CI: 0.004–0.351, q = 0.029). However, these configurations were identified in only 16 scanned repositories, limiting the precision and generalizability of the estimate. None of the remaining five artifacts showed a statistically significant association after multiple-testing correction.

\begin{table*}[!t]
\caption{Firth-adjusted associations between repository artifacts and known-vulnerable dependencies.}
\label{tab:12}
\centering
\scriptsize
\setlength{\tabcolsep}{3pt}
\renewcommand{\arraystretch}{1.12}
\begin{tabularx}{\textwidth}{@{}>{\raggedright\arraybackslash}X>{\raggedright\arraybackslash}X>{\raggedright\arraybackslash}X>{\raggedright\arraybackslash}X>{\raggedright\arraybackslash}X>{\raggedright\arraybackslash}X@{}}
\toprule
\textbf{Artifact} & \textbf{Present (n)} & \textbf{OR} & \textbf{95\% CI} & \textbf{p} & \textbf{q} \\
\midrule
Dependency-update automation configuration & 16 & 0.037 & 0.004-0.351 & 0.004 & 0.029 \\
CI configuration & 103 & 0.435 & 0.164-1.153 & 0.094 & 0.289 \\
Security policy file & 35 & 0.455 & 0.112-1.841 & 0.269 & 0.461 \\
Container build file & 77 & 1.635 & 0.530-5.047 & 0.393 & 0.461 \\
CODEOWNERS file & 11 & 0.079 & 0.003-2.007 & 0.124 & 0.289 \\
Pre-commit configuration & 9 & 1.414 & 0.145-13.799 & 0.766 & 0.766 \\
Any of the six artifacts (composite) & 164 & 0.703 & 0.312-1.585 & 0.395 & 0.461 \\
\bottomrule
\end{tabularx}
\vspace{4pt}
\begin{minipage}{\textwidth}\scriptsize n = 494 for each model. Each artifact was modeled separately, adjusting for log(1 + package count), owner type, log(1 + stars), and grouped construction stratum. OR < 1 indicates lower odds of at least one detected vulnerable dependency. Wald 95\% CIs and Benjamini-Hochberg-adjusted q-values are shown; these sparse exploratory models are not causal.\end{minipage}
\end{table*}

\section{Discussion}

\subsection{Principal Findings}

The principal findings reveal that assessing software supply-chain security in construction-related applications involves challenges beyond identifying dependencies with known vulnerabilities. Two concerns emerges:  incomplete visibility into software dependencies and uncertainty about the security implications of detected vulnerabilities. Only 54.6\% of the studied repositories completed the primary vulnerability scan, while 80.0\% of successfully scanned repositories contained at least one dependency version associated with a published vulnerability advisory. Moreover, vulnerability exposure was not evenly distributed: affected repositories varied substantially in their advisory burden, yet a relatively small number of upstream package families were shared across a large proportion of them. These findings highlight the importance of considering assessment coverage, vulnerability applicability, and shared dependencies when interpreting software supply-chain exposure in construction-related software.

\textbf{Dependency assessability varies across construction subdomains, creating uneven visibility into potential vulnerabilities}. Although the same assessment protocol was applied across the cohort, supported dependency inputs were identified in 67.9\% of construction project-management repositories compared with only 21.4\% of robotics and automation repositories. This difference indicates that a uniform scanning procedure does not necessarily produce comparable coverage across construction functions. Assessability depends partly on the dependency information available in each repository and its compatibility with the selected scanning tools. In particular, construction-related robotics and automation software may use specialized package formats and build systems that fall outside the supported inputs of conventional SCA workflows. Consequently, lower detected vulnerability prevalence in a subdomain cannot be interpreted independently of its assessment coverage. This finding does not imply that less assessable repositories are necessarily more or less secure rather it identifies a visibility gap that must be filled to improve dependency security. For construction software evaluation, scan results should therefore be accompanied by information about supported dependency formats, successfully analyzed components, and unresolved assessment gaps.

\textbf{Known-vulnerable dependencies are widespread among assessable repositories}. The finding that 80.0\% of successfully scanned repositories contained at least one advisory-matched dependency is consistent with previous research documenting vulnerable dependencies in general-purpose software. Prana et al. \cite{ref39} investigated vulnerable dependencies across 450 open-source Java, Python, and Ruby projects, while a commercial audit of 965 codebases reported open-source vulnerabilities in 86\% and high- or critical-risk vulnerabilities in 81\% \cite{ref7}. However, differences in sampling, dependency resolution, vulnerability databases, and assessment procedures prevent direct comparisons of prevalence estimates. Our results, therefore, establish that known-vulnerable dependencies are also prevalent in the assessable portion of construction-related open-source software, without demonstrating whether construction software is more or less exposed than software in other domains.

\textbf{The observed prevalence also depends on the dependency evidence available for assessment}. Repositories analyzed using lockfile inputs exhibited a higher affected proportion (96.2\%) than the primary scanned population (80.0\%). Although this difference is consistent with the possibility that more complete dependency information reveals additional advisory matches, the lockfile subset also differs in repository and ecosystem composition. The analysis therefore cannot attribute the difference to input completeness alone. Furthermore, the primary scan included development dependencies, some of which may not be included in deployed applications. Prior research by Pashchenko et al. \cite{ref36} highlights the importance of distinguishing dependency vulnerabilities identified through static analysis from vulnerabilities relevant to deployed software. These considerations show why advisory counts must be interpreted in relation to dependency scope and application context. A repository with more detected vulnerabilities could reflect broader dependency visibility rather than necessarily poorer security.

Severity and weakness classifications help characterize exposure but provide limited evidence of operational consequences. Nearly all affected repositories (97.2\%) contained at least one \textit{High} or \textit{Critical} severity advisory cluster. Although this indicates widespread exposure to known severe vulnerabilities, it also limits the usefulness of maximum severity as a standalone prioritization criterion: a High- or Critical-severity threshold distinguishes only a relatively small number of affected repositories. Moreover, CVSS base scores describe the technical severity of reported vulnerabilities rather than their consequences in a particular deployment \cite{ref15}. Determining practical relevance requires additional evidence about whether affected components are included in deployed software, whether vulnerable functionality is reachable, and how the affected components are used.

CWE classifications provide another major perspective by identifying the weakness types associated with reported vulnerabilities. Resource-consumption and resource-allocation weaknesses had the broadest repository reach, while path traversal, information exposure, and code injection were also widely represented. These categories can guide more targeted investigation of construction software. For instance, path-traversal advisories may motivate examination of file-processing operations involving BIM models and project documents, while resource-consumption weaknesses may warrant availability-oriented assessment of model-processing services, project dashboards, and site-monitoring applications. Similarly, code-injection and information-exposure advisories may motivate review of data-processing interfaces and access controls. These examples represent potential security review priorities rather than attack paths demonstrated by this study. Because CWE labels were obtained from advisory metadata, their distribution reflects reported weakness classifications rather than independently verified behavior in construction applications. Nevertheless, the observed categories provide a useful starting point for moving from broad severity-based triage toward weakness-specific investigation.

\textbf{Shared upstream dependencies extend known-vulnerability exposure to downstream}. While affected repositories contained a median of 50 unique advisory clusters, vulnerability burden varied substantially across the scanned population. At the same time, a relatively small number of affected package families had broad repository reach: the two most widely shared families collectively appeared with advisory matches in approximately 70\% of affected repositories. These families spanned multiple package ecosystems and supported common software functions, indicating that vulnerability exposure was not limited to construction-specific components. Instead, construction-related applications also inherited known vulnerabilities through general-purpose libraries reused across different development contexts. This shared exposure has implications for organizing vulnerability assessment and remediation efforts. When multiple applications depend on the same affected upstream package family, advisory investigation and identification of available fixes may be coordinated rather than performed independently for every repository. However, cumulative repository reach should not be interpreted as the proportion of vulnerabilities that could be eliminated by updating those packages. Individual repositories may contain additional vulnerable dependencies, and appropriate remediation depends on affected versions, dependency relationships, available fixes, and compatibility constraints. 

\textbf{Observable repository characteristics do not necessarily reflect effective dependency security maintenance}. The repository artifacts represented aspects of development infrastructure, security configuration, and dependency management, but their collective presence was not significantly associated with detected vulnerable dependency exposure. Dependency update automation was the only individual artifact exhibiting a statistically significant negative association. The absence of statistically significant associations for other artifacts does not establish that the corresponding practices are ineffective. The findings instead reinforce the distinction between observable repository configurations and demonstrated security maintenance. For construction software assessment, such artifacts will provide useful contextual information and help to improve the security of dependencies in construction software.

 The findings demonstrate that interpreting known-vulnerability exposure in construction-related software requires further study. Assessment coverage determines which dependencies are observable, dependency scope and application context influence the relevance of detected vulnerabilities, and shared upstream components connect exposure across independently developed repositories. The study contributes empirical evidence on these dimensions across construction software subdomains, showing both the extent of detectable known-vulnerability exposure and the boundaries of conclusions supported by repository-level SCA. These findings provide a basis for more transparent dependency assessment, context-aware vulnerability triage, and coordinated investigation of widely reused affected components.

\subsection{Theoretical Contributions}

The study makes two contributions to construction management theory. The first extends the conception of construction as a loosely coupled system to the software layer of digital construction. Dubois \& Gadde \cite{ref13} located tight coupling within projects and loose coupling in the permanent network of firms. The present results point to a technical layer of coupling that cuts across both. In a cohort spanning six construction functions, ten general-purpose package families, none specific to construction, reached nine in ten affected repositories. Tools built by unrelated developers for different construction functions are thus coupled through shared upstream code: a newly disclosed vulnerability in a widely used package can simultaneously expose every tool that includes an affected version, a pattern resembling the rapid propagation that Perrow \cite{ref37} associated with tight coupling. The buffering that loose coupling ordinarily provides, by sealing off a breakdown in one part of a system \cite{ref47}, does not intercept exposure that arrives through a shared dependency.

We term this condition coupling asymmetry: exposure is tightly coupled across the sector, while the capacity to respond remains loosely coupled within temporary multi-organizations that lack durable channels for coordinated action \cite{ref9}. As a proposition for future testing, coupling asymmetry suggests why construction software risk may not be manageable at the project level alone, the level at which project cyber-risk assessments are typically conducted \cite{ref50,ref49}. A project team can assess its own tools, but the leverage for reducing exposure lies in a small number of upstream components maintained outside any construction organization. The concept also clarifies the dual meaning of the concentration result. Concentration is at once the source of correlated exposure and the basis for coordinated, sector-level remediation, and which of the two prevails depends on whether the loosely coupled industry can act on shared components collectively.

The second contribution is the construct of software-composition assessability. Construction cyber-risk frameworks identify assets, threats, and risk factors \cite{ref46,ref50}, but they implicitly assume that the components of a digital tool can be known. In this cohort, the assumption could not be verified with the selected protocol for 45.4\% of construction-related repositories, and 78.6\% of robotics and automation repositories lacked a supported dependency input. Assessability is therefore a precondition construct: risk cannot be assessed, prioritized, or assigned for components that cannot be observed.

Separating assessability from exposure also exposes a common inferential error. An absent alert may mean that no vulnerable component exists, or that no component could be seen, and only an explicit assessability record distinguishes the two. We propose that construction cyber-risk frameworks treat assessability as a distinct dimension, measured before and alongside vulnerability exposure. Doing so answers the call of Orton \& Weick \cite{ref31} to specify the elements and mechanisms of coupling: in digital construction, the coupled elements are software components, the mechanism is dependency reuse, and assessability determines whether the coupling is visible to those who must manage it.

\subsection{Practical Implications}

For clients, contractors, and technology suppliers, the findings translate into three actions. First, composition transparency should be a condition of acceptance for digital deliverables. Because the absence of an alert is uninterpretable without evidence of what was inspected, a delivered tool should be accompanied by an SBOM that meets recognized minimum elements \cite{ref29}, a vulnerability scan tied to the delivered version, and a record of any components that could not be assessed. This extends the security-minded information management of ISO 19650-5:2020 from project information to the software that processes it. Transparency should be required most insistently where assessability is lowest, particularly for robotics, automation, and embedded site systems.

Second, triage should be coordinated at the level of shared components rather than individual alerts. Affected repositories carried a median of 50 advisory clusters, so forwarding raw alerts to project teams is unlikely to be effective. Because ten package families reached 90.4\% of affected repositories, a research center, software supplier, or contractor maintaining several tools can assess a recurring family once, determine whether its vulnerable code is shipped and reachable \cite{ref38}, and reuse that assessment across tools. Construction functions such as model interpretation, quantity takeoff, or scheduling logic still require regression testing after an update. Prioritization should combine reach with severity and the weakness type recorded for the advisory, for example by asking whether untrusted external files reach an affected parser in a BIM exchange service, or whether an affected web component handles externally accessible requests in a project platform.

Third, maintenance ownership should be assigned in a way that survives project handover. The association between dependency-update automation and lower exposure is consistent with evidence that automated security updates are frequently accepted when maintainers are active \cite{ref2}, although the present data cannot separate the effect of automation from that of active maintenance. Automation is therefore best specified together with a named maintainer and a process for handling new advisories after acceptance, consistent with the NIST Secure Software Development Framework \cite{ref44}. These practices are timely. The EU Cyber Resilience Act’s reporting obligations apply from 11 September 2026, and its remaining obligations from 11 December 2027 \cite{ref5} hence, technology suppliers serving EU construction markets, and the contractors who integrate their products, will therefore need the component-level evidence that this study found was often unavailable.

\section{Conclusion, Limitations, and Future Research Directions}

This study characterized dependency assessment coverage and known-vulnerability advisory matches in 904 public construction-related GitHub repositories. Of these, 494 completed a scan under the selected protocol, and 395 contained at least one matched package version. Among affected repositories, 97.2\% had a maximum scanner-derived severity classified as \textit{High} or \textit{Critical}. CWE labels described recurring resource-consumption, path-traversal, information-exposure, and injection weaknesses, while ten general-purpose package families collectively reached 90.4\% of affected repositories. These findings describe the assessed repository inputs and the vulnerability information available during scanning. They do not establish industry prevalence, deployment, exploitability, or operational consequences.

The contribution is a construction-defined baseline that distinguishes assessment coverage from detected findings and identifies upstream components recurring across public construction-related tools. The observed shared-component pattern motivates coupling asymmetry as a proposition for future research on maintenance coordination; organizational response capacity was not measured here. For prospective users and maintainers, the practical implication is to evaluate software functionality together with evidence of its composition, the applicability of known advisories, and responsibility for continued maintenance.

Several limitations bound these conclusions. First, the cohort is a relevance-ranked, non-probability sample of public, non-fork GitHub repositories retrieved through 26 queries, with at most one page of 100 results per query. Search covered names, descriptions, and topics rather than README text or source-code contents. The resulting coverage is incomplete and cannot be quantified as a recall rate. Screening established documented construction relevance and executable content, not production maturity or deployment by construction organizations. Low star counts and user ownership do not establish poor quality, but neither do the eligibility criteria establish operational adoption. Shared owners, templates, and upstream components may also make repository observations dependent. The reported intervals do not account for all such dependence and should not be interpreted as supporting population-wide generalization.

Second, two reviewers independently assessed repository eligibility and assigned each eligible repository to a primary construction stratum. Agreement was strong, and all disagreements were resolved through discussion until 100\% consensus was reached for the final analytic classifications. Nevertheless, both eligibility and stratum assignment involved human judgment. Interpretations of repository purpose, executable evidence, and multifunctionality may therefore be subjective and may contain residual classification error. The resulting strata should be understood as transparent operational classifications rather than infallible ground truth.

Third, assessment depended on one scanner, its configuration and supported formats, and the vulnerability information available during scanning. No independent validation established detection precision or recall, and different tools or databases may produce different findings. A completed scan established that at least one package was parsed, not that the repository’s full composition was recovered. Repositories without supported inputs may have had no external dependencies, incomplete declarations, or unsupported dependency formats; the analysis did not distinguish these explanations. Where manifest processing relied on external resolution, identified versions may reflect scan-time metadata rather than a historical build environment. Preserving the input files therefore does not guarantee identical results from a later rescan.

Fourth, the primary outcome included development and runtime dependencies where represented in the assessed inputs. The path-based sensitivity analysis excluded selected input locations but did not identify runtime-only package exposure. Deployment scope, vulnerable-code reachability, and exploitation were not established. Severity summaries used scanner-reported numeric maxima and consequently emphasize the highest available values rather than a harmonized comparison of scoring systems or deployment risk. Native CWE labels inherited variation in advisory quality and abstraction level, included overlapping hierarchical entries, and were unavailable for 167 clusters. These measures support descriptive characterization and follow-up review, not direct estimates of construction-site harm.

Fifth, the cross-sectional design does not establish when maintainers became aware of a vulnerability, whether a fix was available to them, or how long remediation took. Collection-time matches can include advisories published after the repository commit and should not be interpreted as evidence that maintainers knowingly retained vulnerable dependencies. The artifact models were exploratory, used sparse indicators and Wald intervals, and did not adjust for ecosystem membership or input type. Their associations may therefore reflect measurement and population differences as well as maintenance practices. Finally, no matched non-construction cohort was included, so the analysis cannot identify a construction-specific excess or deficit of advisory matches.

Future research should first validate repository findings against actual construction deployments, including installed versions, dependency scope, and reachable functionality. Independent scanners and manual assessment could help characterize detection disagreement and identify reasons for incomplete coverage. Longitudinal studies should then distinguish changes in repository inputs, resolved dependencies, and advisory information, rather than interpreting a single assessment as evidence of maintenance delay.

Further work should investigate assessment coverage for robotics, embedded systems, and construction-specific software formats, and compare construction-related repositories with appropriately matched non-construction samples. Organizational studies are also needed to test coupling asymmetry by examining whether shared dependencies are accompanied by fragmented responsibility, duplicated assessment effort, or delayed remediation across project and supplier boundaries.

\section*{Declaration of Competing Interest}

The authors declare no known competing financial interests or personal relationships that could have influenced the work reported in this article.

\section*{Funding}

This research did not receive any specific grant from funding agencies in the public, commercial, or not-for-profit sectors.

\section*{Data Availability}

The supplementary appendices report the screening codebook and search queries (Appendix A), the construction strata definitions, screening counts, and inter-rater agreement (Appendix A), the dependency-evidence taxonomy by ecosystem (Appendix B), the severity coding thresholds (Appendix C), the provenance record and SHA-256 manifest hashes (Appendix D), and the software inventory and reproduction command sequence (Appendix E). The frozen commit list, selected dependency inputs, aggregate scanner outputs, CWE enrichment outputs, and analysis code are available from the corresponding author on reasonable request. Repository-specific findings that could identify exploitable components are not publicly ranked and will be shared only through a responsible-disclosure process.

\section*{Ethics and Responsible Disclosure}

All data came from public repositories and public vulnerability databases; no private communications or contributor-level personal data were collected. The study did not recruit human participants and did not execute repository code. Findings are reported in aggregate, and repository-specific actionable rankings are not included in the manuscript.

\FloatBarrier

\end{document}